\documentclass[journal]{IEEEtran}
\usepackage{graphicx}
\usepackage{url}
\usepackage{algorithm,algorithmic}
\usepackage{multirow}
\usepackage{color}

\usepackage{cite}
\ifCLASSINFOpdf
\else
\fi
\usepackage{amsmath}
\usepackage{amsfonts}
\begin{document}
\title{Generating Watermarked Adversarial Texts}

\author{Mingjie Li, Hanzhou Wu and Xinpeng Zhang
\thanks{\emph{Corresponding author: Dr. Hanzhou Wu, Email: h.wu.phd@ieee.org}}
}


\markboth{}%
{}



\maketitle

\begin{abstract}
Adversarial example generation has been a hot spot in recent years because it can cause deep neural networks (DNNs) to misclassify the generated adversarial examples, which reveals the vulnerability of DNNs, motivating us to find good solutions to improve the robustness of DNN models. Due to the extensiveness and high liquidity of natural language over the social networks, various natural language based adversarial attack algorithms have been proposed in the literature. These algorithms generate adversarial text examples with high semantic quality. However, the generated adversarial text examples may be maliciously or illegally used. In order to tackle with this problem, we present a general framework for generating watermarked adversarial text examples. For each word in a given text, a set of candidate words are determined to ensure that all the words in the set can be used to either carry secret bits or facilitate the construction of adversarial example. By applying a word-level adversarial text generation algorithm, the watermarked adversarial text example can be finally generated. Experiments show that the adversarial text examples generated by the proposed method not only successfully fool advanced DNN models, but also carry a watermark that can effectively verify the ownership and trace the source of the adversarial examples. Moreover, the watermark can still survive after attacked with adversarial example generation algorithms, which has shown the applicability and superiority.
\end{abstract}


\IEEEpeerreviewmaketitle

\section{Introduction}
\IEEEPARstart{A}{lthough} deep neural networks (DNNs) have achieved great success in many tasks such as computer vision \cite{google:googlenet, olaf:unet, kaiming:resnet} and natural language processing \cite{rico:NMT, wu:NMT, EMNLP:dChen}, they are vulnerable to adversarial examples \cite{arXiv:intriguing, arXiv:explaining}. Adversarial examples are inputs (to DNN models) intentionally designed by an attacker to cause a target model to make incorrect outputs. They can be generated by applying a hardly perceptible perturbation to the original samples. From the perspective of defense, adversarial examples can be used for improving the robustness of DNNs, e.g., by mixing adversarial examples into the training set \cite{arXiv:universal}.

Many advanced adversarial example generation (AEG) algorithms \cite{arXiv:intriguing, arXiv:explaining, arXiv:AlexeyKurakin:Adversarial, EuroSP:NicolasPapernot:TheLimitations, IEEE:SeyedMohsenMoosaviDezfooli:DeepFool} have been proposed in the past years. For example, digital image has been widely adopted for AEG since the majority of DNNs are originally designed for visual tasks which require the input data to be images. Mainstream image based AEG algorithms produce adversarial images by modifying the pixels. The adversarial image example will not introduce noticeable artifacts since the modification degree is highly slight. Generating speech adversarial examples has also attracted increasing interest because of the wide application of speech recognition in daily life. For example, by inserting a well-designed noise to a speech signal, the speech adversarial example can cause a target speech recognition system to output any specified sentence \cite{IEEE:NCarlini:audio}. Some representative speech AEG algorithms can be found in references \cite{USENIX:NicholasCarlini:Hidden, arXiv:YaoQin:Imperceptible, arXiv:MoustafaAlzantot:DidYouHearThat, arXiv:YuanGong:CraftingAdversarial}.

In natural language processing (NLP), due to the widespread use of DNNs, there are also increasing concerns about the security of NLP systems, among which research on \emph{adversarial natural language (text) examples} has become more and more important. However, due to the discrete nature of text content, gradient perturbation \cite{arXiv:intriguing, arXiv:explaining} and other AEG algorithms originally designed to images and speech signals cannot be directly applied to texts. In terms of human perception, it is difficult to understand perturbed texts while slight modifications to image pixels can still yield meaningful images. Therefore, generating adversarial texts is a very challenging topic. Mainstream text based AEG algorithms can be divided to character-level adversarial attacks \cite{NAACL:SteffenEger:TextProcessing, ACL:JavidEbrahimi:HotFlip, arXiv:YonatanBelinkov:SyntheticAndNatural}, word-level adversarial attacks \cite{ACL:YuanZang:WordlevelTextual, EMNLP:MoustafaAlzantot:GeneratingNatural, arXiv:NicolasPapernot:CraftingAdversarial} and sentence-level adversarial attacks \cite{ACL:MarcoTulioRibeiro:Semantically, ComputerNetworks:Jialiang:Asentence, arXiv:MohitIyyer:AdversarialExample}. Among them, word-level AEG models perform comparatively well on both attack efficiency and adversarial text quality \cite{ACL:YuanZang:WordlevelTextual, arXiv:XiaosenWang:NaturalLanguage}.

Adversarial examples can be crafted to cause a target DNN model to misclassify. On the other hand, as mentioned above, adversarial examples can be also used for constructing robust DNN models. For example, they can be employed during DNN training in order to supply new training data from which the model might benefit \cite{LREC:UsingAdversarialExamples}. In this sense, adversarial examples should be protected. As AEG models become more and more mature, it will surely raise concerns that adversarial examples may be maliciously or illegally used. It is therefore necessary to seek solutions to protect adversarial examples.

The aforementioned analysis has motivated us to study how to protect adversarial examples. In particular, we will focus on protecting adversarial texts in this paper. A straightforward idea is to build sophisticated access control protocols to protect adversarial texts, which, however, has limited control after the adversarial texts have been shared with authorized users. An alternative option is \emph{digital watermarking} \cite{book:digitalWatermarking}, which embeds secret data into a cover signal by slightly modifying the noisy component of the cover \cite{Wu:TCSVT2017, Wu:TCSVT2021}. By extracting secret data from the watermarked signal, we can identify the ownership and trace the source of the watermarked signal. Obviously, it is quite desirable to use digital watermarking for adversarial texts as long as the watermarked adversarial texts well maintain the attack efficiency and adversarial text quality.

One may use traditional text watermarking algorithms \cite{SPIE:Alattar:WatermarkingElectronic, IEEE:Li:NovelTextWatermarking, IEEE:Chen:TextWatermarking, generation:paper1, generation:paper2} directly to mark adversarial texts, which, however, is not suitable for practice. The reason is, generating adversarial texts and watermarking texts are two independent task. Performing well on one task does not mean it results in good performance on the other task. For example, after embedding a secret watermark, the watermarked ``adversarial text'' may be no longer adversarial. On the other hand, comparing with watermarking,  ``adversarial'' watermarking requires more redundancy of the original text (which is often highly encoded), which makes it more difficult to embed a watermark. Therefore, watermarking adversarial texts is an important yet very challenging task.

In this paper, we propose a general framework to embed a secret watermark into a given text. In the proposed work, given the original text that is not adversarial, for each word to be probably modified, a set of candidate words are determined to ensure that all the words in the set can be used to replace the present word to carry secret data or construct the adversarial text. By applying a word-level adversarial text generation algorithm, the generated text is not only adversarial, but also carries a watermark that can verify the ownership and trace the source of the adversarial text. Experimental results have demonstrated the superiority and applicability.

The rest structure of this paper is organized as follows. First of all, we provide the preliminary concepts in Section II. The proposed framework is then detailed in Section III. Thereafter, we conduct convinced experiments and also provide analysis in Section IV. Finally, we conclude this paper in Section V.

\section{Preliminaries}
In this section, we briefly review the advances in adversarial text generation and text watermarking so that we can better introduce the proposed work in the subsequent section.

\subsection{Adversarial Text Generation}
Increasing adversarial text generation models, ranging from character-level \cite{NAACL:SteffenEger:TextProcessing, ACL:JavidEbrahimi:HotFlip, arXiv:YonatanBelinkov:SyntheticAndNatural} to sentence-level \cite{ACL:MarcoTulioRibeiro:Semantically, ComputerNetworks:Jialiang:Asentence, arXiv:MohitIyyer:AdversarialExample}, are proposed in recent years. Character-level models usually generate adversarial texts by modifying individual characters in the original texts. Though character-level adversarial text generation models have been proven to be effective, they may cause the perceptibility problem because changing individual characters often leads to invalid words, which can be easily recognized by humans. Moreover, character-level adversarial attacks are easy to be defended, e.g., preprocessing the input text with a spell checking tool can resist against such attacks \cite{arXiv:ZhaoMeng:AGeometry}. Sentence-level attacks can be realized by generating new sentences such as inserting an additional sentence to the original text \cite{EMNLP:RobinJia:AdversarialExamples}, and rewriting the original sentence by an encoder-decoder model \cite{arXiv:MohitIyyer:AdversarialExample}. For sentence-level adversarial attacks, a most important requirement is that the new sentence should match the original text semantically.

Different from character-level/sentence-level attacks, word-level attacks generate adversarial text examples by replacing some individual words with new words in the original texts, resulting in new texts that have the same semantics to the original ones. On the one hand, synonym replacement keeps the semantic distortion between the adversarial text and the original text within a low level. On the other hand, since only a part of words are to be replaced, the computational complexity is low as well by applying an efficient optimization algorithm. In addition, it has been demonstrated that word-level attacks achieve superior performance in fooling target models \cite{ACL:YuanZang:WordlevelTextual}, motivating us to focus on word-level attacks in this paper.

The goal of a word-level adversarial text generation model is to craft adversarial texts using a limited vocabulary that can successfully fool the victim model(s), which actually requires us to solve a combinatorial optimization problem. The implementation of word-level adversarial text generation models is generally divided into two key steps: 1) reducing search space, and; 2) searching for adversarial texts \cite{ACL:YuanZang:WordlevelTextual}. In this subsection, we briefly introduce  three typical word-level adversarial text generation models, which are ``Embedding/LM+Genetic'' \cite{EMNLP:MoustafaAlzantot:GeneratingNatural}, ``Synonym+Greedy'' \cite{ACL:ShuhuaiRen:GeneratingNatural}, and ``Sememe+PSO'' \cite{ACL:YuanZang:WordlevelTextual}. These models will be used in our experiments.

\subsubsection{Embedding/LM+Genetic} Alzantot \emph{et al}. \cite{EMNLP:MoustafaAlzantot:GeneratingNatural} use a population-based black box optimization algorithm to generate adversarial texts with semantics and syntax similar to the original texts, which are capable of fooling well-trained victim models. In the algorithm, the restriction on vocabulary embedding distance and the restriction on language model prediction score are combined to reduce the search space. For searching suitable words, a popular metaheuristic population evolution algorithm is used. The model generates adversarial texts on sentiment analysis and textual entailment datasets. Experimental results show that they achieve good performance in attacking well-trained victim models.

\subsubsection{Synonym+Greedy} Research on adversarial text generation models is mainly based on word substitution strategies, avoiding artificial manufacturing and realizing automatic generation. Ren \emph{et al}. \cite{ACL:ShuhuaiRen:GeneratingNatural} propose a new word replacement order determined by word saliency and classification probability based on the \emph{synonyms} \cite{ACM:GAMiller:WordNet}  replacement strategy, and propose a probability-weighted word saliency greedy algorithm to form the adversarial text generation model. Experiments show that the generated adversarial texts greatly reduce the accuracy of the text classification model under the condition of using low replacement rates that are difficult for humans to perceive.

\subsubsection{Sememe+PSO} Due to inappropriate word space reduction methods and undesirable adversarial text search methods, word-level adversarial text generation models can be further improved. By combining the sememe-based word substitution method with the particle swarm optimization based search algorithm, Zang \emph{et al.} \cite{ACL:YuanZang:WordlevelTextual} propose a new word-level adversarial text generation model. Through experiments, it is known that the model obtains the higher attack success rates, and the generated adversarial texts are of the higher quality.

\subsection{Text Watermarking}
Text watermarking (or called \emph{linguistic watermarking}, \emph{natural language watermarking}) can be modeled as a communication problem. A data encoder embeds secret information in a cover text by slightly modifying the cover. The resulting text containing hidden information (also called \emph{marked text}) should be sent to a data decoder and may be attacked during transmission. After receiving the probably attacked and marked text, the data decoder is able to reliably extract the embedded information for ownership verification or other purposes.

Text watermarking is often evaluated by the rate-distortion performance. For a payload, the distortion between the cover text and the marked text should be as low as possible. The term ``distortion'' often measures the semantic difference between the cover text and the marked text. On the other hand, when the distortion is kept within a fixed level, it is desirable to embed information as much as possible. In addition, since the marked text may be altered by an adversary, it is further required that a text watermarking technique should be robust to malicious attacks so that secret information can be still extracted from the attacked marked text for ownership protection.

Text watermarking can be realized by modifying text format \cite{IEEE:NurulShamimiKamaruddin:AReview}, e.g., secret bits can be embedded by adjusting the white space between two words or lines \cite{SPIE:Alattar:WatermarkingElectronic}. However, these format based algorithms have limited robustness since the text format can be easily changed by an intentional attacker, resulting in a high bit error rate. Mainstream algorithms exploit the syntactic and semantic nature of texts for watermark embedding \cite{SCN:Kim:Adaptive, ACM:Kim:TextWatermarking, IH:Atallah:Natural, IEEE:Mali:Implementation}. They emphasize that the critical semantic information of the marked texts are consistent with the original ones. For example, secret bits can be embedded into a text by replacing specified words in the text with semantically-similar words. In this paper, we will study text watermarking along this direction to adapt to the construction of adversarial texts.

\begin{figure}[!t]
\centering
\includegraphics[width=\linewidth]{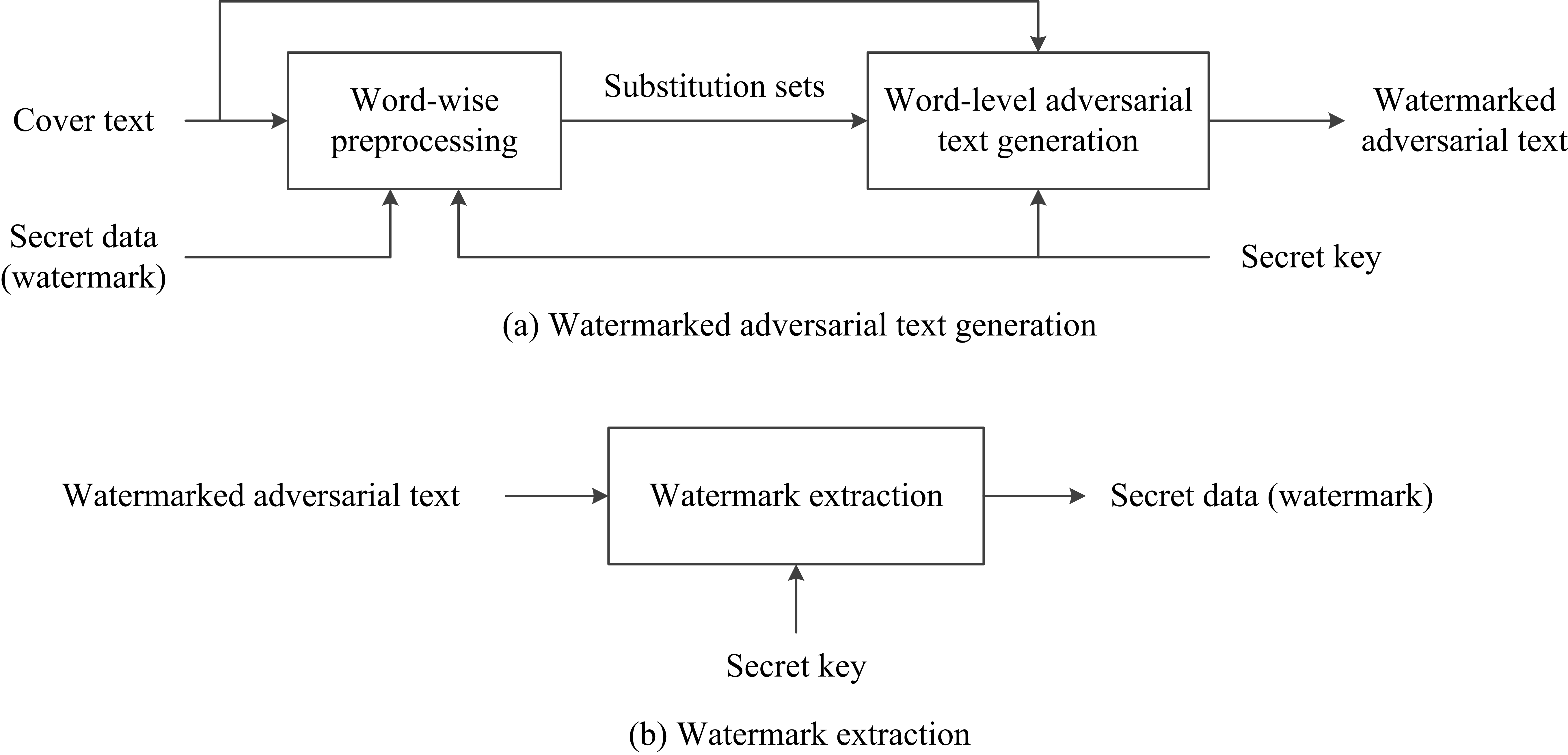}
\caption{Sketch for the proposed adversarial text watermarking system.}
\end{figure}

\section{Proposed Method}
\subsection{Overview}
As shown in Fig. 1, the proposed work embeds a secret watermark during generating the adversarial text, which considers text watermarking and adversarial text generation as a whole and achieves superior performance in text watermarking and adversarial text generation. For data embedding, the proposed framework first processes the cover text (i.e., the original text) to construct a substitution set for each word in the text. The substitution set of a word in the specific position of the cover text includes an indefinite number of candidate words that can be used to replace the original word in order to construct the watermarked adversarial text. In this way, according to the secret key and the substitution sets, the secret watermark can be successfully embedded into the cover text through a word-level adversarial text generation procedure. The resulting new text not only fools the specific DNN model, but also carries a secret watermark identifying its ownership. With the secret key, the secret watermark can be easily reconstructed from the watermarked adversarial text, without the need of constructing the substitution sets. In the following, we show more details.

\subsection{Adversarial Text Watermarking}
Given a cover text $\textbf{x} = \{x_1, x_2, ..., x_n\}$, $n\geq 1$, the proposed watermarked adversarial text generation procedure produces a watermarked adversarial text $\textbf{y} = \{y_1, y_2, ..., y_n\}$, from which a secret watermark $\textbf{w} = \{w_1, w_2, ..., w_m\}\subset \{0,1\}^m$, $m \geq 1$, can be retrieved according to a secret key. To achieve this goal, we first preprocess the cover text and then apply a word-level adversarial text generation algorithm to the cover text. 

\begin{algorithm}[!t]
 \caption{Word-wise processing procedure}
 \begin{algorithmic}[1]
	\renewcommand{\algorithmicrequire}{\textbf{Input:}}
	\renewcommand{\algorithmicensure}{\textbf{Output:}}
	\REQUIRE Cover text $\textbf{x}$, secret watermark $\textbf{w}$.
	\ENSURE Substitution sets $S(x_1), S(x_2), ..., S(x_n)$.
	\STATE Determine $I_1, I_2, I_3$ (using Algorithm 2, described latter)
	\FOR{each $i\in I_1$}
	\STATE Set $S(x_i) = \{x_i\}$
	\ENDFOR
	\FOR{each $i\in I_2$}
	\STATE Determine $T(x_i)$ from a vocabulary
	\STATE Set $S(x_i) = T(x_i)$
	\ENDFOR
	\FOR{each $i\in I_3$}
	\STATE Determine $T(x_i)$ from a vocabulary
	\STATE Determine $T_0(x_i), T_1(x_i), ..., T_{2^l-1}(x_i)$ using $f$
	\STATE Determine the present secret binary stream $\textbf{w}_i$ (that has a length of $l$) to be embedded from $\textbf{w}$
	\STATE Set $S(x_i) = T_{d(\textbf{w}_i)}(x_i)$
	\ENDFOR
	\RETURN $S(x_1), S(x_2), ..., S(x_n)$
 \end{algorithmic}
\end{algorithm}

\subsubsection{Word-wise Preprocessing}
The proposed word-wise preprocessing procedure aims to construct a substitution set $S(x_i)$ for each $x_i\in\textbf{x}$. All words in $S(x_i)$ are semantically close to $x_i$ so that $x_i$ can be changed to $y_i\in S(x_i)$ in the subsequent word-level adversarial text generation procedure. It is possible that $y_i = x_i$, i.e., $x_i$ can be unchanged. Let $I=\{1, 2, ..., n\}$ be an index set. Thus, each word $x_i\in \textbf{x}$ can be uniquely indexed by $i\in I$. It is possible for two different $i\neq j$ that $x_i = x_j$, i.e., $x_i$ and $x_j$ are corresponding to the same word. To construct $S(x_1), S(x_2), ..., S(x_n)$, $I$ is partitioned into three disjoint subsets $I_1$, $I_2$ and $I_3$. In other words, we always have $I_1\cap I_2=\emptyset$, $I_1\cap I_3=\emptyset$, $I_2\cap I_3=\emptyset$ and $I_1\cup I_2\cup I_3=I$.

Assuming that, $I_1$, $I_2$ and $I_3$ have been previously determined. The words corresponding to $I_1$ will be unchanged. It indicates that $S(x_i) = \{x_i\}$ for all $i\in I_1$. For each $i\in I_2\cup I_3$, a list of candidate words $T(x_i) = \{t_{i,1}, t_{i,2}, ..., t_{i,n_i}\}$ can be collected from a large-scale vocabulary so that each word in $T(x_i)$ is semantically consistent with $x_i$. For example, one may determine $T($``see''$)$ = \{``see'', ``look'', ``watch''\}. For each $i\in I_2$, we set $S(x_i) = T(x_i)$. For each $i\in I_3$, we determine $S(x_i)$ based on $T(x_i)$ and the secret data to be embedded.

Specifically, for each $i\in I_3$, we use a mapping function $f$ to map each word $e\in T(x_i)$ to a binary string that has a length of $l > 0$. In this way, $T(x_i)$ can be partitioned into a total of $2^l$ disjoint subsets according to the mapped binary strings. Let $T_0(x_i), T_1(x_i), ..., T_{2^l-1}(x_i)$ be the $2^l$ subsets. $T_v(x_i)$ contains all words that are mapped to a binary string whose decimal value is $v$, e.g., if $l = 3$ and $f($``see''$)$ = $(101)_2$, ``see'' will be an element of $T_5($``see''$)$ since $((101)_2)_{10} = 5$. Obviously, we have $T_a(x_i)\cap T_b(x_i) = \emptyset$ for any $0\leq a < b\leq 2^l-1$ and $T(x_i) = \cup_{k=0}^{2^l-1}T_k(x_i)$. Only those words (in $\textbf{x}$) corresponding to $I_3$ will be used to carry the secret data. Assuming that we need to embed a binary stream $\textbf{w}_i$ (that has a length of $l$) into some $x_i$ $(i\in I_3)$, we set $S(x_i) = T_{d(\textbf{w}_i)}(x_i)$, where $d(\textbf{w}_i)$ is the decimal value of $\textbf{w}_i$, e.g., $d($``011''$) = 3$. Algorithm 1 shows the pseudo-code to construct $S(x_1), S(x_2), ..., S(x_n)$.

\subsubsection{Word-level Adversarial Text Generation}
Once the substitution sets are determined, we are able to generate $\textbf{y}$ through an efficient word substitution based adversarial text generation algorithm. That is, the adversarial text generation algorithm is to select exactly one word from each $S(x_i)$, $i\in I$, to construct $\textbf{y}$, so that $\textbf{y}$ is an adversarial text example. In other words, we have $y_i\in S(x_i)$ for all $1\leq i\leq n$. Since all words in $S(x_i)$ ($i\in I_3$) are mapped to the same secret binary stream, each $y_i$ ($i\in I_3$) surely carries a secret binary stream, which indicates that, $\textbf{y}$ is also marked after text generation. 

\subsubsection{Watermark Extraction}
In order to verify the ownership, all words corresponding to $I_3$ carrying secret binary streams can be collected from the adversarial text example. Then, all the secret binary streams can be extracted from these words and further concatenated to form the entire watermark $\textbf{w}$ that can be used for ownership verification.

\emph{Remark 1:} $l$ is pre-determined according to the secret key. A smaller $l$ for a fixed $I_3$ indicates that the maximum number of embeddable bits is smaller, but the text quality may be better since more candidate words can be used. A larger $l$ provides a larger embedding capacity, but may cause some substitution sets to be empty that may lead to failed embedding. Following previous methods that use a word to carry at most one bit, we will also use $l = 1$ throughout this paper. 
% $l < 1$ means that $1/l$ words are used to carry one bit, e.g., we can use a word-pair to carry one bit, meaning that $l = 1/2$. In this case, we can consider multiple words as a whole in order to construct the corresponding substitution sets.

\emph{Remark 2:} It is free for us to design the mapping function $f$. For simplicity, we use the Hash function MD5\footnote{https://en.wikipedia.org/wiki/MD5} to construct $f$. In detail, given a word, we first compute its MD5 value and then use the $l$ least bits to denote the mapped binary stream. Obviously, one may also use a secret key to decide what bits should be taken out to constitute the mapped binary stream.

\begin{table*}[!t]
\centering
     \caption{Examples of applying the proposed method (with ``Sememe+PSO'') to attack BiLSTM on the IMDB dataset. The modified words in the watermarked adversarial texts are highlighted with underlines. ``bpw'' means ``bits per word''.}
\begin{tabular}{|c|c|l|}

\hline
\begin{tabular}[c]{@{}c@{}}Original text\\ (prediction = ``\emph{negative}'')
\end{tabular}                  & \multicolumn{2}{l|}{\begin{tabular}[c]{@{}l@{}}\emph{This movie was perhaps the biggest waste of 2 hours of my life. From the opening 10 minutes, I was ready to}\\ \emph{leave. The cliches there slapping you in the face, and the plot was not only predictably stupid, but full of more}\\ \emph{holes than swiss cheese. I am considering suing for that lost 2 hours, and \$6.25 along with the fact that I am}\\ \emph{now stupider for watching this waste of film. The T-Rex's must be flipping in their graves, so to speak.}\end{tabular}}                                         \\ \hline
\multirow{4}{*}{\begin{tabular}[c]{@{}c@{}}\\ \\ \\ \\ \\ \\ \\ \\Watermarked adversarial texts\\ (prediction = ``\emph{positive}'')\end{tabular}} & 0.05 bpw                                                     & \begin{tabular}[c]{@{}l@{}}
\emph{This movie was perhaps the biggest waste of 2 hours of my life. From the opening 10 minutes,}\\
\emph{I was ready to leave. The cliches there slapping you in the face, and the plot was not only}\\ \emph{predictably stupid,but full of more holes than swiss cheese. I am \underline{\textbf{rapprochement}} suing for that}\\ \emph{lost 2 hours, and \$6.25 along with the fact that I am now stupider for watching this waste of}\\ \emph{film. The T-Rex's must be flipping in their graves, so to speak.}\end{tabular}  \\ \cline{2-3} 
& 0.10 bpw  & 
\begin{tabular}[c]{@{}l@{}}\emph{This movie was perhaps the biggest waste of 2 hours of my life. From the opening 10 minutes,}\\ \emph{ I was ready to leave. The cliches there slapping you in the face, and the plot was \underline{\textbf{either}} only}\\ \emph{predictably stupid, but full of more holes than swiss cheese. I am \underline{\textbf{rapprochement}} suing for that}\\ \emph{lost 2 hours, and \$6.25 along with the fact that I am now stupider for watching this waste of}\\ \emph{film. The T-Rex's must be flipping in their graves, so to speak.}\end{tabular} \\ \cline{2-3} 
& 0.15 bpw  & 
\begin{tabular}[c]{@{}l@{}}\emph{This \underline{\textbf{film}} was perhaps the biggest waste of 2 \underline{\textbf{yorker}} of my life. From the opening 10 minutes,}\\ \emph{ I was ready to leave. The cliches there \underline{\textbf{civilian}} you in the face, and the \underline{\textbf{clown}} was not only}\\ \emph{predictably stupid, but full of more holes than swiss cheese. I am considering suing for that}\\ \emph{lost 2 hours, and \$6.25 along with the fact that I am now stupider for watching this waste of}\\ \emph{film. The T-Rex's must be flipping in their graves, so to speak.}\end{tabular}       \\ \cline{2-3} 
& 0.20 bpw  & 
\begin{tabular}[c]{@{}l@{}}\emph{This movie was perhaps the biggest waste of 2 \underline{\textbf{yorker}} of my life. From the opening 10 minutes,}\\ \emph{I was ready to leave. The cliches there \underline{\textbf{civilian}} you in the face, and the \underline{\textbf{clown}} was not only}\\ \emph{predictably stupid, but full of more holes than swiss cheese. I am \underline{\textbf{rapprochement}} suing for that}\\ \emph{lost 2 hours, and \$6.25 along with the \underline{\textbf{sub}} that I am now stupider for watching this waste of}\\ \emph{film. The T-Rex's must be flipping in their graves, so to speak.}\end{tabular} \\ \hline
\end{tabular}
\end{table*}

\begin{table*}[!t]
\centering
     \caption{Examples of applying the proposed method (with ``Sememe+PSO'') to attack BiLSTM on the SST dataset.}
\begin{tabular}{|c|c|l|}

\hline
\begin{tabular}[c]{@{}c@{}}Original text\\  (prediction = ``\emph{negative}'')\end{tabular}                  & \multicolumn{2}{l|}{\begin{tabular}[c]{@{}l@{}}
\emph{Sunshine state lacks the kind of dynamic that limbo offers, and in some ways is a rather indulgent piece.}\end{tabular}}\\ \hline
\multirow{4}{*}{\begin{tabular}[c]{@{}c@{}} \\Watermarked adversarial texts\\  (prediction = ``\emph{positive}'')\end{tabular}}  & 0.05 bpw & \begin{tabular}[c]{@{}l@{}}
 \emph{Sunshine state lacks the kind of dynamic that limbo offers, and in some ways is a rather}\\ \emph{\underline{\textbf{unbridled}} piece.}\end{tabular}  \\ \cline{2-3} & 0.10 bpw  & \begin{tabular}[c]{@{}l@{}}
 \emph{Sunshine state lacks the kind of dynamic that limbo offers, and in some ways is a \underline{\textbf{passably}}}\\ \emph{indulgent piece.}\end{tabular} \\ \cline{2-3} & 0.15 bpw & \begin{tabular}[c]{@{}l@{}}
 \emph{Sunshine state \underline{\textbf{needs}} the kind of dynamic that limbo offers, and in some ways is a rather}\\ \emph{\underline{\textbf{unbridled}} piece.}\end{tabular} \\ \cline{2-3} & 0.20 bpw  & \begin{tabular}[c]{@{}l@{}}
 \emph{\underline{\textbf{Moon}} \underline{\textbf{reason}} lacks the kind of dynamic that limbo offers, and in some ways is a \underline{\textbf{reason}}}\\ \emph{indulgent piece.} \end{tabular} \\ \hline
\end{tabular}
\end{table*} 

\subsection{Construction of Disjoint Subsets}
It is open for us to determine $I_1$, $I_2$ and $I_3$ in advance as long as $I_3$ can fully carry the watermark information and the data receiver can determine $I_3$ for extracting the watermark. For example, one may use a secret key to randomly produce $I_1$, $I_2$ and $I_3$, which, however, does not take advantage of the subsequent word-level adversarial text generation algorithm and therefore may limit the adversarial text quality since word-level adversarial text generation algorithms often require us to take into account the part of speech of the word to be modified to generate adversarial texts with high quality. 

To this end, we present an efficient strategy for constructing $I_1$, $I_2$ and $I_3$. In detail, we first determine the part of speech of each word in $\textbf{x}$, and then add the index of each word to the corresponding set according to the part of speech of the word. For example, assuming that there are at most $n_p > 0$ parts of speech in a text, we therefore initialize $n_p$ index sets $P_1$, $P_2$, ..., $P_{n_p}$ as empty for $\textbf{x}$, e.g., one may use $P_1$ to include all indexes of words (in $\textbf{x}$) that are nouns, $P_2$ to include all indexes of words (in $\textbf{x}$) that are adjectives and so on. Then, for each $x_i\in \textbf{x}$, $1\leq i\leq n$, we determine the part of speech of $x_i$ and then add $i$ to the corresponding index set $P_{s(i)}$, where $s(i)$ means the index of the part of speech set of $x_i$. All words corresponding to $P_{s(i)}$ have the same part of speech. It can be inferred that, $P_j\cap P_k=\emptyset$ for all $j\neq k$ and $\cup_{j=1}^{n_p}P_j = I$. 

Suppose that, the used word-level adversarial text generation algorithm $\mathcal{A}$ only allows us to modify words with specific parts of speech, e.g., only modifying verbs and nouns. 
 Let $J\subset I$, we define $Q(J) = \cup_{j\in J}P_{s(j)}$. Obviously, $Q(I) = I$. 
Without the loss of generalization, there must exist exactly one $J\subset I$ that $Q(J)$ corresponds to all modifiable words in $\textbf{x}$ given $\mathcal{A}$. It is inferred that $|Q(J)| \leq |I| = n$, where $|*|$ means the size of a set. Thus, we can set $I_1 = I\setminus Q(J)$ and $I_2\cup I_3 = Q(J)$. Let $L$ be the bit-length of the secret watermark. If we use each word to carry exactly $l$ bits, a secret key can be used to randomly select $L/l$ elements from $Q(J)$ to constitute $I_3$. The rest elements will be then used to constitute $I_2$. In this way, the three disjoint subsets are successfully constructed. From the implementation of view, Algorithm 2 shows the pseudo-code to construct $I_1$, $I_2$ and $I_3$, from which we can find that the time complexity is linear with respect to $n$. 

\begin{algorithm}[!t]
 \caption{Construction of disjoint subsets}
 \begin{algorithmic}[1]
	\renewcommand{\algorithmicrequire}{\textbf{Input:}}
	\renewcommand{\algorithmicensure}{\textbf{Output:}}
	\REQUIRE Cover text $\textbf{x}$, text generation algorithm $\mathcal{A}$, secret key, the bit-length of the secret watermark $L$, parameter $l$.
	\ENSURE Disjoint subsets $I_1$, $I_2$, $I_3$.
	\STATE Initialize $J = I_1 = I_2 = I_3 = \emptyset$ and $I = \{1, 2, ..., n\}$
	\FOR{each $i\in I$}
	\IF{$x_i\in \textbf{x}$ is modifiable according to $\mathcal{A}$}
	\STATE Set $J = J\cup \{i\}$
	\ELSE
    \STATE Set $I_1 = I_1\cup \{i\}$
	\ENDIF
	\ENDFOR
	\STATE Randomly select $L/l$ elements out from $J$ to constitute $I_3$ according to the secret key (notice that $L/l \leq |J|$)
	\STATE Set $I_2 = I_3\setminus J$
	\RETURN $I_1$, $I_2$, $I_3$
 \end{algorithmic}
\end{algorithm}

\section{Experimental Results and Analysis}
In this section, we are to conduct extensive experiments and analysis to evaluate the proposed work.

\subsection{Datasets and Models}
We choose two popular benchmark datasets, i.e., IMDB \cite{IMDB:paper} and SST-2 \cite{SST2:paper}, to evaluate the performance of the proposed method. The IMDB dataset includes 25,000 training texts and 25,000 testing texts for movie reviews, labeled as positive and negative. The SST-2 dataset includes 6,920 training texts, 872 validation texts and 1,821 testing texts. Both datasets are commonly used in binary sentiment classification. However, the average sentence length of the texts in the SST-2 dataset is much shorter than that in the IMDB dataset. Therefore, it is more difficult to generate adversarial texts on SST-2. 
We adopt two widely used sentence encoding models bidirectional LSTM (BiLSTM) \cite{BiLSTM:paper} and BERT \cite{BERT:paper} as the victim models. In our simulation, the hidden states of the BiLSTM is 128-dimensional and the 300-dimensional pre-training GloVe \cite{GloVe:paper} was used for word embedding. Additionally, we use three above-mentioned adversarial text generation models: Embedding/LM+Genetic \cite{EMNLP:MoustafaAlzantot:GeneratingNatural}, Synonym+Greedy \cite{ACL:ShuhuaiRen:GeneratingNatural}, Sememe+PSO \cite{ACL:YuanZang:WordlevelTextual}, for generating the watermarked adversarial texts. Obviously, these algorithms can be used to generate the corresponding adversarial texts without any hidden information, i.e., the adversarial texts are non-watermarked.

\begin{table*}[!t]
     \centering
     \caption{The attack success rates (\%) of the proposed method based on different attack models with different watermark payload sizes when attacking the victim models BiLSTM and BERT.}
\begin{tabular}{c||c||c||c||cccc}
\hline\hline
\multirow{2}{*}{\begin{tabular}[c]{@{}c@{}}Victim  Model\end{tabular}} & \multirow{2}{*}{Dataset} & \multirow{2}{*}{Attack Model} & \multirow{2}{*}{Non-watermarked} & \multicolumn{4}{c}{Watermarked} \\
    &    &  &   & 0.05 bpw  & 0.10 bpw   & 0.15 bpw  & 0.20 bpw    \\ \hline
\multirow{6}{*}{BiLSTM}                                                 & \multirow{3}{*}{IMDB}     & Embedding/LM+Genetic          & 87.30                 & 86.90   & 86.50   & 71.70   & 69.10   \\
                                                                        &                           & Synonym+Greedy               & 90.20                 & 70.10   & 52.60   & 38.40   & 38.60   \\
                                                                        &                           & Sememe+PSO                   & 100.00                & 88.40   & 82.70   & 64.50   & 55.20   \\ \cline{2-8} 
                                                                        &                           & Embedding/LM+Genetic          & 69.10                 & 53.30   & 55.60   & 50.90   & 45.20   \\
                                                                        & SST-2                     & Synonym+Greedy               & 70.20                 & 60.90   & 58.60   & 53.20   & 50.40   \\
                                                                        &                           & Sememe+PSO                   & 89.80                 & 75.70   & 55.40   & 51.30   & 50.70   \\ \hline
\multirow{6}{*}{BERT}                                                   & \multirow{3}{*}{IMDB}     & Embedding/LM+Genetic          & 88.40                 & 82.10   & 85.20   & 71.80   & 68.40   \\
                                                                        &                           & Synonym+Greedy               & 70.60                 & 60.60   & 46.90   & 34.70   & 30.50   \\
                                                                        &                           & Sememe+PSO                   & 94.50                 & 86.30   & 79.30   & 58.70   & 48.50   \\ \cline{2-8} 
                                                                        & \multirow{3}{*}{SST-2}    & Embedding/LM+Genetic          & 60.10                 & 53.50   & 46.20   & 40.50   & 36.10   \\
                                                                        &                           & Synonym+Greedy               & 62.80                 & 59.70   & 58.10   & 49.40   & 48.20   \\
                                                                        &                           & Sememe+PSO                   & 85.60                 & 73.10   & 52.90   & 47.60   & 48.10 \\ \hline\hline
\end{tabular}
\end{table*}

\begin{table*}[!t]
     \centering
     \caption{The mean modification rates (\%) of the watermarked adversarial texts generated by the proposed method based on different attack models with different watermark payload sizes when attacking the victim models BiLSTM and BERT.}
\begin{tabular}{c||c||c||c||cccc}
\hline\hline
\multirow{2}{*}{\begin{tabular}[c]{@{}c@{}}Victim Model\end{tabular}} & \multirow{2}{*}{Dataset} & \multirow{2}{*}{Attack Model} & \multirow{2}{*}{Non-watermarked} & \multicolumn{4}{c}{Watermarked} \\
    &  &  &   & 0.05 bpw   & 0.10 bpw & 0.15 bpw  & 0.20 bpw \\ \hline
\multirow{6}{*}{BiLSTM}                                                 & \multirow{3}{*}{IMDB}    & Embedding/LM+Genetic                & 7.01                 & 7.14   & 9.26   & 12.03  & 12.97 \\
                                                                        &                          & Synonym+Greedy                & 6.59                 & 6.77   & 6.61   & 9.62   & 10.08 \\
                                                                        &                          & Sememe+PSO                    & 3.32                 & 3.50   & 5.05   & 8.26   & 11.14 \\ \cline{2-8} 
                                                                        &                          & Embedding/LM+Genetic           & 10.28                & 11.99  & 11.31  & 12.37  & 13.32 \\
                                                                        & SST-2                    & Synonym+Greedy                & 9.08                 & 10.18  & 11.70  & 12.23  & 13.18 \\
                                                                        &                          & Sememe+PSO                    & 9.29                 & 9.32   & 9.74   & 12.21  & 13.10 \\ \hline
\multirow{6}{*}{BERT}                                                   & \multirow{3}{*}{IMDB}    & Embedding/LM+Genetic           & 7.07                 & 7.09   & 7.13   & 9.08   & 10.15 \\
                                                                        &                          & Synonym+Greedy                & 3.82                 & 4.76   & 6.22   & 7.34   & 8.82  \\
                                                                        &                          & Sememe+PSO                    & 3.31                 & 3.34   & 4.75   & 7.26   & 10.06 \\ \cline{2-8} 
                                                                        & \multirow{3}{*}{SST-2}   & Embedding/LM+Genetic           & 9.27                 & 9.35   & 10.06  & 10.04  & 11.06 \\
                                                                        &                          & Synonym+Greedy                & 8.12                 & 8.14   & 9.20   & 9.88   & 10.75 \\
                                                                        &                          & Sememe+PSO                    & 7.76                 & 8.02   & 9.16   & 8.34   & 12.03 \\ \hline\hline
\end{tabular}
\end{table*}

\subsection{Basic Settings and Evaluation Metrics}
In experiments, we empirically set the the population size to 60 and the maximum number of iterations to 20 for the two population-based models ``Embedding/LM+Genetic'' and ``Sememe+PSO''. The other hyper-parameters for the three word-level adversarial text generation models are consistent with the recommended hyper-parameters in the original papers. For training BiLSTM, we set the epoch to 20 and the batch size to 64. We use TensorFlow for simulation and a single NVIDIA TITAN RTX 24GB GPU for accelerating model training.

We evaluate the proposed method in terms of two aspects, i.e., adversarial text attack performance and text watermarking performance if not otherwise specified. The adversarial attack performance is measured on the attack success rates on victim models. The attack success rate is defined as the percentage of attack attempts that make the victim model output the target label. The watermarking performance is mainly evaluated by the quality of the generated watermarked adversarial texts, the watermark payload and the watermark detection accuracy. We use three metrics to evaluate the quality of the watermarked adversarial texts, i.e., modification rate, grammaticality, and fluency. The modification rate is defined as the percentage of the modified words in the watermarked adversarial text. We measure the grammaticality by counting the increase rate of grammatical errors in the watermarked adversarial text (compared with the original text) with the LanguageTool\footnote{https://languagetool.org/}. The fluency of each watermarked adversarial text will be measured by the language model \emph{perplexity (PPL)} with the help of GPT-2 \cite{GPT2:paper}. Notice that, both adversarial attack and watermarking require that the generated texts should have good quality.

\subsection{Performance Evaluation on Adversarial Attack}
Due to the limited computational resource and large search space for adversarial text construction, for each experiment, we randomly select 500 correctly classified texts from the corresponding testing set to generate watermarked adversarial texts for evaluation. During experiments, we limit the length of each original text to the range [10, 120] for efficient evaluation. 

We set the watermark payload size to the range [0.05, 0.2] to ensure that both the attack success rate and the quality of the watermarked adversarial text are high. Here, the payload size is defined as the ratio between the number of watermark bits and the length of the original text, i.e., $L/n$. The corresponding \emph{non-watermarked} adversarial texts are generated as well under the same experimental condition for fair comparison.

Table I and Table II show examples of using the proposed method for text generation. In Table I and Table II, ``0.05 bpw'' represents the payload size, i.e., each word carries 0.05 bits. Here, ``bpw'' is short for bits per word. From Table I and Table II, we can find that the original text can be modified with a few words to become an adversarial text, which verifies the superiority of the used AEG algorithm. On the other hand, we can find that the word-modification rates become higher when the payload size becomes larger, which is reasonable because a larger payload size means that more words are likely to be modified in order to carry more secret bits. 

The attack success rates of the proposed method based on different attack models are shown in Table III. Comparing with the non-watermarked adversarial texts (i.e., the payload size is 0), the success rates for the watermarked adversarial texts are relatively lower. We observe that with the increase of the watermark payload size, the attack success rates of the proposed method based on different attack models gradually decrease when attacking the victim models on different datasets. The attack success rates for IMDB are generally higher than that for SST-2 under the same conditions. It reveals that it is more difficult to generate watermarked adversarial texts on SST-2, which is due to the shorter average length of the text samples. We can easily find that the attack success rates of the proposed method based on different attack models can achieve higher than 30\% in all cases when attacking the victim models, which reveals the vulnerability of DNNs and demonstrates the superiority of the AEG procedure. It also indicates that the proposed watermarking strategy does not significantly impair the attack performance. 

\begin{table*}[!t]
     \centering
     \caption{The average increase rates of grammatical errors (\%) of the non-watermarked adversarial texts and the watermarked adversarial texts under different experimental conditions.}
\begin{tabular}{c||c||c||c||cccc}
\hline\hline
\multirow{2}{*}{\begin{tabular}[c]{@{}c@{}}Victim Model\end{tabular}} & \multirow{2}{*}{Dataset} & \multirow{2}{*}{Attack Model} & 
\multirow{2}{*}{Non-watermarked} & \multicolumn{4}{c}{Watermarked} \\
    &   &   &  & 0.05 bpw   & 0.10 bpw   & 0.15 bpw  & 0.20 bpw \\ \hline
\multirow{6}{*}{BiLSTM}                                                 & \multirow{3}{*}{IMDB}    & Embedding/LM+Genetic           & 6.21                 & 6.60    & 7.85   & 8.31   & 8.36  \\
                                                                        &                          & Synonym+Greedy                & 4.96                 & 5.07    & 5.01   & 6.27   & 7.43  \\
                                                                        &                          & Sememe+PSO                    & 1.82                 & 2.03    & 2.05   & 2.96   & 3.06  \\ \cline{2-8} 
                                                                        &                          & Embedding/LM+Genetic           & 6.88                 & 7.14    & 7.03   & 8.89   & 9.13  \\
                                                                        & SST-2                    & Synonym+Greedy                & 4.98                 & 5.15    & 5.82   & 6.90   & 7.22  \\
                                                                        &                          & Sememe+PSO                    & 4.10                 & 4.31    & 4.86   & 4.39   & 4.97  \\ \hline
\multirow{6}{*}{BERT}                                                   & \multirow{3}{*}{IMDB}    & Embedding/LM+Genetic           & 5.65                 & 5.73    & 6.98   & 7.19   & 7.68  \\
                                                                        &                          & Synonym+Greedy                & 4.23                 & 4.35    & 4.79   & 5.02   & 6.47  \\
                                                                        &                          & Sememe+PSO                    & 1.75                 & 2.13    & 2.16   & 2.73   & 2.85  \\ \cline{2-8} 
                                                                        & \multirow{3}{*}{SST-2}   & Embedding/LM+Genetic           & 6.17                 & 6.26    & 7.01   & 7.36  & 8.72  \\
                                                                        &                          & Synonym+Greedy                & 4.52                 & 4.81    & 5.93  & 6.48   & 6.51  \\
                                                                        &                          & Sememe+PSO                    & 3.26                 & 3.28    & 4.20   & 3.79   & 4.65  \\ \hline\hline
\end{tabular}
\end{table*}

\begin{table*}[!t]
     \centering
     \caption{The mean language model perplexity (PPL) for non-watermarked adversarial texts and watermarked adversarial texts under different experimental conditions.}
\begin{tabular}{c||c||c||c||cccc}
\hline\hline
\multirow{2}{*}{\begin{tabular}[c]{@{}c@{}}Victim Model\end{tabular}} & \multirow{2}{*}{Dataset} & \multirow{2}{*}{Attack Model} & \multirow{2}{*}{Non-watermarked} & \multicolumn{4}{c}{Watermarked} \\
  & & &  & 0.05 bpw & 0.10 bpw & 0.15 bpw & 0.20 bpw  \\ \hline
\multirow{6}{*}{BiLSTM}                                                 & \multirow{3}{*}{IMDB}     & Embedding/LM+Genetic          & 112.41  & 113.50  & 129.86 & 150.96 & 154.57 \\
                                                                        &                           & Synonym+Greedy               & 103.57  & 111.83  & 109.49 & 139.45 & 143.04 \\
                                                                        &                           & Sememe+PSO                   & 81.25   & 99.28   & 101.26 & 130.17 & 135.43 \\ \cline{2-8} 
                                                                        &                           & Embedding/LM+Genetic          & 287.12  & 345.16  & 315.23 & 380.51 & 472.55 \\
                                                                        & SST-2                     & Synonym+Greedy               & 281.63  & 308.63  & 355.32 & 376.57 & 406.29 \\
                                                                        &                           & Sememe+PSO                   & 256.58  & 316.14  & 323.65 & 349.28 & 375.40  \\ \hline
\multirow{6}{*}{BERT}                                                   & \multirow{3}{*}{IMDB}     & Embedding/LM+Genetic          & 95.17   & 106.51  & 113.27 & 130.24 & 145.16 \\
                                                                        &                           & Synonym+Greedy               & 93.54   & 109.30  & 120.65 & 125.13 & 141.67 \\
                                                                        &                           & Sememe+PSO                   & 80.49   & 98.14  & 105.42 & 121.58 & 127.25 \\ \cline{2-8} 
                                                                        & \multirow{3}{*}{SST-2}    & Embedding/LM+Genetic          & 288.62  & 310.06 & 336.14 & 358.60 & 454.38  \\
                                                                        &                           & Synonym+Greedy               & 280.70  & 307.35 & 323.61 & 342.93 & 356.32 \\
                                                                        &                           & Sememe+PSO                   & 247.58  & 295.67 & 308.15 & 300.70 & 322.07 \\ \hline\hline
\end{tabular}
\end{table*}

\begin{table*}[!t]
\centering
     \caption{The mean accuracy (\%) of watermark extraction after attacking watermarked adversarial texts.}
\scalebox{0.9}{
\begin{tabular}{c||c||c||cccccccccccc}
\hline\hline
\multirow{4}{*}{\begin{tabular}[c]{@{}c@{}}Victim Model\end{tabular}} & \multirow{4}{*}{Dataset} & \multirow{4}{*}{Watermark Attacking Method} & \multicolumn{12}{c}{Proposed Method}       \\ \cline{4-15}  &  &   & \multicolumn{4}{c|}{Embedding/LM+Genetic}  & \multicolumn{4}{c|}{Synonym+Greedy}  & \multicolumn{4}{c}{Sememe+PSO} \\ \cline{4-15}  & & & \multicolumn{4}{c|}{Payload (bpw)} & \multicolumn{4}{c|}{Payload (bpw)} & \multicolumn{4}{c}{Payload (bpw)} \\
 &  &  & 0.05  & 0.10   & 0.15  & \multicolumn{1}{c|}{0.20}   & 0.05  & 0.10   & 0.15  & \multicolumn{1}{c|}{0.20}   & 0.05     & 0.10     & 0.15    & 0.20    \\ \hline
\multirow{6}{*}{BiLSTM} & \multirow{3}{*}{IMDB} & Embedding/LM+Genetic & 97.51 & 93.90 & 92.91 & \multicolumn{1}{c|}{92.46} & 98.16 & 97.93 & 94.97 & \multicolumn{1}{c|}{92.70} & 98.53    & 96.56   & 91.43   & 90.40  \\ &  & Synonym+Greedy & 98.68 & 94.22 & 93.29 & \multicolumn{1}{c|}{92.87} & 98.02 & 97.81 & 93.55 & \multicolumn{1}{c|}{92.64} & 99.71  & 97.26  & 92.08  & 91.62  \\ &  & Sememe+PSO & 97.71 & 95.17 & 93.34 & \multicolumn{1}{c|}{94.41} & 99.20 & 98.82 & 96.13 & \multicolumn{1}{c|}{93.14} & 98.42    & 96.23   & 90.23   & 87.30  \\ \cline{2-15} & \multirow{3}{*}{SST-2}   & Embedding/LM+Genetic & 96.83 & 93.87 & 92.80 & \multicolumn{1}{c|}{92.32} & 97.61 & 97.50 & 92.68 & \multicolumn{1}{c|}{92.02} & 98.40    & 94.31   & 90.67   & 90.28  \\ &              & Synonym+Greedy   & 96.86 & 94.73 & 92.93 & \multicolumn{1}{c|}{92.47} & 97.12 & 98.73 & 93.50 & \multicolumn{1}{c|}{91.93} & 98.44    & 96.24   & 93.83   & 91.51  \\ &  & Sememe+PSO & 97.05 & 95.11 & 91.22 & \multicolumn{1}{c|}{93.26} & 98.70 & 97.81 & 94.77 & \multicolumn{1}{c|}{92.84} & 98.27    & 94.15   & 89.65   & 90.04  \\ \hline \multirow{6}{*}{BERT}  & \multirow{3}{*}{IMDB}    & Embedding/LM+Genetic & 98.46 & 96.01 & 93.95 & \multicolumn{1}{c|}{92.87} & 99.83 & 97.61 & 94.66 & \multicolumn{1}{c|}{92.80} & 99.94    & 96.72   & 93.93   & 91.76  \\ & & Synonym+Greedy & 98.71 & 96.62 & 94.04 & \multicolumn{1}{c|}{92.93} & 98.89 & 97.10 & 94.54 & \multicolumn{1}{c|}{92.68} & 99.91    & 98.80   & 93.72   & 90.83  \\ & & Sememe+PSO  & 98.82 & 97.30 & 94.57 & \multicolumn{1}{c|}{93.26} & 98.90 & 98.94 & 92.71 & \multicolumn{1}{c|}{93.91} & 98.86    & 96.48   & 93.05   & 88.43  \\ \cline{2-15}  & \multirow{3}{*}{SST-2}   & Embedding/LM+Genetic           & 97.36 & 95.73 & 94.05 & \multicolumn{1}{c|}{92.34} & 98.72 & 97.67 & 94.28 & \multicolumn{1}{c|}{91.71} & 99.57    & 96.45   & 92.17   & 91.72  \\
 &   & Synonym+Greedy  & 97.55 & 95.94 & 96.01 & \multicolumn{1}{c|}{92.42} & 98.17 & 96.80 & 94.50 & \multicolumn{1}{c|}{92.53} & 99.70    & 97.27   & 92.96   & 91.81  \\     &   & Sememe+PSO    & 97.80 & 96.26 & 94.47 & \multicolumn{1}{c|}{93.60} & 99.85 & 97.91 & 95.05 & \multicolumn{1}{c|}{92.98} & 98.98    & 95.94   & 91.64   & 90.35  \\ \hline\hline
\end{tabular}}
\end{table*}

\subsection{Performance Evaluation on Text Watermarking}
The (mean) modification rates for non-watermarked adversarial texts and watermarked adversarial texts with different watermark payload sizes are shown in Table IV. Comparing with the non-watermarked adversarial texts, the modification rates for the watermarked adversarial texts are relatively higher. Moreover, the modification rates increase as the watermark payload size becomes larger. The modification rates for the watermarked adversarial texts generated on IMDB are generally lower than that on SST-2 under the same conditions. This is also mainly due to the fact that the average length of texts of IMDB is much longer than that of SST-2. Overall, the modification rates are all less than 14\% and the maximum difference between the (mean) modification rate for watermarked texts and that for non-watermarked texts is less than 6\%, indicating that the modification degree can be kept within a low level. It means that the proposed framework does not significantly modify the original text, which implies that the adversarial text quality after watermarking can be maintained.

The average increase rates of grammatical errors of the non-watermarked adversarial texts and the watermarked adversarial texts generated by the proposed method are shown in Table V. Comparing with the non-watermarked adversarial texts, the average increase rates of grammatical errors of the watermarked adversarial texts are relatively higher. And we can clearly see that the average increase rates of grammatical errors of the watermarked adversarial texts gradually increase with the increase of the payload size, and the average increase rates of grammatical errors for adversarial texts generated on the IMDB dataset are generally lower than that on the SST-2 dataset. It is related to the fact that the modification rates increase with the increase of the watermark payload size, and the modification rates on the IMDB dataset are generally lower than that on the SST-2 dataset.

We use PPL to measure the fluency of the generated adversarial texts. It is generally believed that the larger the PPL, the worse the quality of the corresponding text given the original text. The mean PPLs for non-watermarked adversarial texts and watermarked adversarial texts under different conditions are shown in Table VI. Comparing with the non-watermarked adversarial texts, the (mean) PPLs for the watermarked adversarial texts are generally higher. It is seen that the PPLs gradually increase with the increase of the watermark payload size, meaning that the text quality gradually declines. The PPLs tested on the SST-2 dataset are generally higher than that tested on the IMDB dataset. Overall, the (mean) PPLs difference between the watermarked adversarial texts and the non-watermarked adversarial texts can be kept low for low payload sizes, which means that the text quality is satisfactory.

In practice, an attacker may attempt to remove the hidden watermark by modifying a watermarked adversarial text. It is reasonably assumed that the modification should not impair the adversarial characteristics of the marked text. Therefore, it is necessary for the proposed method to resist against such kind of modification, which is referred to as robustness. To evaluate the robustness, we mimic a real-world attack scenario that is: \emph{given a watermarked adversarial text, the attacker applies an AEG algorithm to the watermarked adversarial text so that the watermark can be removed while the text is still adversarial}. 

In Subsection IV-C, we have generated 500 watermarked adversarial texts for each experiment. We use the three word-level AEG algorithms to attack these watermarked adversarial texts. After attacking, these texts are still adversarial, but the watermark is distorted. The distortion can be measured by the percentage of correctly reconstructed watermark bits, which is defined as \emph{accuracy}. Obviously, a higher accuracy means that more bits are correctly recovered, which reveals better robustness. Table VII shows the mean (watermark) accuracy after attacking watermarked adversarial texts with different watermark attacking methods. From Table VII, we can find that the accuracy reaches higher than 90\% for all cases though it gradually decreases with the increase of the payload size. 

Moreover, in most cases, the watermark attacking method used to attack the watermarked adversarial texts based on the watermark attacking method itself makes the mean accuracy lower than the other watermark attacking methods, which is reasonable because different attacking methods use words of different parts of speech for AEG and obviously the watermark attacking method used to attack the watermarked adversarial texts based on the watermark attacking method itself is more likely to erase the watermark information. Nevertheless, overall, the proposed method shows satisfactory robustness.

\section{Conclusion}
In this paper, aiming to protect adversarial text examples from being abused, we propose a novel method to embed a secret watermark into a text while keeping its adversarial attack ability. By comparing the watermarked adversarial texts with the non-watermarked adversarial texts, our experiments show that the proposed method well maintains the adversarial attack ability and achieves satisfactory text quality. In other words, the watermarked adversarial texts generated by the proposed method not only successfully fool the target models but also enable us to retrieve a secret watermark for identifying the ownership and source information. In future, we will improve both the adversarial attack performance and the watermarking performance. Though the proposed work is designed to texts, it can be extended to other media objects. We hope this attempt can make a contribution to the cross field involving AEG and digital watermarking. 

\section*{Acknowledgement}
This work was supported in part by the National Natural Science Foundation of China (NSFC) under Grant 61902235 and Grant U1936214; and in part by the Shanghai ``Chen Guang'' Program under Grant 19CG46; and in part by the Science and Technology Commission of Shanghai Municipality (STCSM) under Grant 21010500200. 

%\ifCLASSOPTIONcaptionsoff
%  \newpage
%\fi

\end{document}